\documentclass[usenatbib]{mnras}
\usepackage{newtxtext,newtxmath}
\usepackage[T1]{fontenc}
\usepackage{ae,aecompl}

\usepackage{graphicx}	
\usepackage{amssymb}
\usepackage{amsmath}	
\usepackage{amssymb}	

\title{A transiting super-Earth close to the inner edge of the habitable zone of an M0 dwarf star}

\author[E.D. Alonso et al.]{E. D\'iez Alonso$^{1}$\thanks{E-mail: diezenrique@uniovi.es},
J. I. Gonz\'alez Hern\'andez$^{2,3}$,
B. Toledo--Padr\'on$^{2,3}$,
\newauthor S. L. Su\'arez G\'omez$^{4}$,
A. Su\'arez Mascare\~no$^{5}$,
D. S. Aguado$^{2}$,
C. Gonz\'alez Guti\'errez$^{1}$,
\newauthor A. Cabrera-Lavers$^{2,6}$,
J. Carballido--Landeira$^{4}$,
L. Bonavera$^{4}$,
F. J. de Cos Juez$^{1}$, 
\newauthor R. Rebolo$^{2,3,7}$ 
\\
$^{1}$Department of Exploitation and Exploration of Mines, University of Oviedo, Oviedo, Spain\\
$^{2}$Instituto de Astrof\'isica de Canarias, E--38205 La Laguna, Tenerife, Spain\\
$^{3}$Universidad de La Laguna, Dpto. Astrof\'isica, E--38206 La Laguna, Tenerife, Spain\\
$^{4}$Departamento de F\'isica, Universidad de Oviedo, C. Federico Garc\'ia Lorca 18, E-33007, Oviedo, Spain\\
$^{5}$Observatoire Astronomique de l\textquotesingle Universit\'e de G\`eneve, 1290 Versoix, Switzerland\\
$^{6}$GRANTECAN, Cuesta de San Jos\'e s/n, E-38712, Bre\~na Baja, La Palma, Spain\\
$^{7}$Consejo Superior de Investigaciones Cient\'ificas, Spain\\
}

\date{Accepted . Received }

\pubyear{2018}

\begin{document}
\label{firstpage}
\pagerange{\pageref{firstpage}--\pageref{lastpage}}
\maketitle

\begin{abstract}
We present a super-Earth orbiting close to the inner edge of the habitable zone of the cool dwarf star K2-286 (EPIC 249889081), detected with data from the K2 mission in its $15^{th}$ campaign. The planet has radius of $2.1\pm0.2$ R$_{\oplus}$, near the 1.5 -- 2.0 R$_{\oplus}$ gap in the radii distribution. The equilibrium temperature is $347^{+21}_{-11}$ K, cooler than most of the small planets with well measured masses, and the orbital period is $27.359\pm0.005$ days. K2-286, located at a distance of $76.3\pm0.3$ pc, is an M0V star with estimated effective temperature of $3926\pm100$ K, less active than other M dwarf stars hosting exoplanets. The expected radial velocity semi-amplitude induced by the planet on the star is $1.9^{+1.3}_{-1.2}$ m$\cdot$s$^{-1}$, and the amplitude of signals in transit transmission spectroscopy is estimated at $5.0\pm3.0$ ppm.
Follow-up observations for mass measurements and transit spectroscopy should be desirable for this relatively bright target ($m_V=12.76, m_{Ks}=9.32$) hosting a transiting super-Earth within the inner edge of the habitable zone.

\end{abstract}

\begin{keywords}
planets and satellites: detection -- techniques: photometric -- techniques: spectroscopic -- stars: low mass -- stars: individual: EPIC 249889081, K2-286
\end{keywords}



\section{INTRODUCTION}

In the search and study of super-Earth and Earth-type planets, low mass stars (0.1 M$_{\odot}$ < M < 0.6 M$_{\odot}$) are primary targets. These stars account for 70\% of the stellar population in the Galaxy \citep{2006AJ....132.2360H,2010AJ....139.2679B}, and planets in such stars tend to be terrestrial as the stellar mass decreases \citep{2012ApJS..201...15H,2013ApJ...767...95D,2015ApJ...807...45D,2015ApJ...798..112M,2015ApJ...814..130M,2018haex.bookE.153M}. Since transiting planets induce deeper dimming and also stronger radial velocity ($RV$) signals, their detection by these methods is easier. The amplitudes of signals in transit transmission spectroscopy \citep{2002ApJ...568..377C} are higher for stars of smaller radius, and planets transiting bright low-mass stars are therefore suitable for extensive atmospheric characterization
\citep{2010Natur.464.1161S,2014Natur.505...69K}.

With the appropriate physical conditions, planets orbiting within the habitable zone of its host star might be able to support liquid water \citep{1993Icar..101..108K}, commonly an assumed condition in the search for extraterrestrial life. Planets orbiting within the habitable zone of M dwarf stars are closer and have shorter orbital periods which favors their detection.
These planets are also excellent targets to study potentially habitable atmospheres \citep{2011ApJ...733...35K,2014ApJ...781...54R}.

The rate of discoveries of potentially habitable planets orbiting low mass stars is monotonically increasing \citep{2016ApJ...817L..20W, 2016Natur.536..437A,2017Natur.542..456G}. However, their habitability is subject of debate; these planets orbit very close to their host star, and experience strong gravitational interactions which result in tidal locking \citep{1964hpfm.book.....D,1993Icar..101..108K}. These stars are generally very active \citep{1998A&A...331..581D,2014ApJ...797..121H}, in particular at early stages in their lifetimes, and the level of activity is usually higher for very late  spectral types \citep{2004AJ....128..426W}. Planets with close - in orbits around M dwarf stars are exposed to strong flare activity and UV and X-ray irradiation from their host star \citep{2014ApJ...797..121H,2018ApJ...860L..30H}.

A large number of transiting planets (2372 confirmed and 2426 candidates to date 2018 November 16)\footnote{Number of candidates and confirmed exoplanets found by Kepler in its first and second missions has been obtained from http://exoplanetarchive.ipac.caltech.edu.}) have been detected by NASA's Kepler mission \citep{2010Sci...327..977B}. The satellite continued observing different ecliptic plane fields in its second mission \footnote{The Kepler spacecraft was retired on 2018 October 30.} \citep{2014PASP..126..398H} with temporal windows spanning $\sim$80 days. Many transiting candidates (355 confirmed and 473 candidates to date 2018 November 16) have been found in K2 data \citep[e.g.][]{2015ApJ...800...59V,2016ApJS..226....7C,2018MNRAS.476L..50D,2018MNRAS.tmpL.105A}.

Campaign 15 focused in the Scorpius region, centered at $\alpha=15:34:28$, $\delta=-20:04:44$, between 2017 August 23 and 2017 November 30. In this campaign, 23,279 targets have been observed at standard long cadence mode and 119 targets at short cadence mode \footnote{https://keplerscience.arc.nasa.gov}. 
	
In this work, we present the detection of a super-Earth transiting the M0V star K2-286 (EPIC 249889081, $\alpha$=15:33:28.7, $\delta$=-16:46:23.72). K2-286 has a Kepler magnitude $K_{p}=12.2$, so K2-286b is a very favorable target to study its atmospheric properties through transit and secondary eclipse measurement due to the increment in the signal quality that a bright star implies.

In section 2 we describe the characterization of the star K2-286 and the analysis of the K2 photometric time series. We also discuss possible contaminating sources, the main parameters derived for the planet, different models for the habitable zone of K2-286 and also estimate the mass of K2-286b. In section 3 we discuss the fact that K2-286b has estimated radius around the upper edge of the 1.5-2.0 R$_{\oplus}$  gap in the radii distribution of exoplanets, and describe its suitability for future characterization by radial velocity and transmission spectroscopy follow-up, comparing K2-286b with other small transiting exoplanets with well measured masses. In section 4 we summarize the main conclusions derived from this work.

\section{METHODS AND DATA ANALYSIS}
\subsection{Stellar characterization}\label{sec:stechar}
Three medium-resolution spectra ($\lambda/\delta \lambda\sim2500$) covering the $UVRI$ bands (spectral ranges 348$-$461 nm, 444$-$604 nm, 571$-$768 nm and 733$-$980 nm, respectively) were obtained for K2-286 with the OSIRIS camera-spectrograph \citep{2000SPIE.4008..623C} of the 10.4 m Gran Telescopio Canarias (GTC), located at Observatorio Roque de los Muchachos in La Palma (Canary Islands, Spain).

\begin{figure*}
	\includegraphics [width=0.9\textwidth]{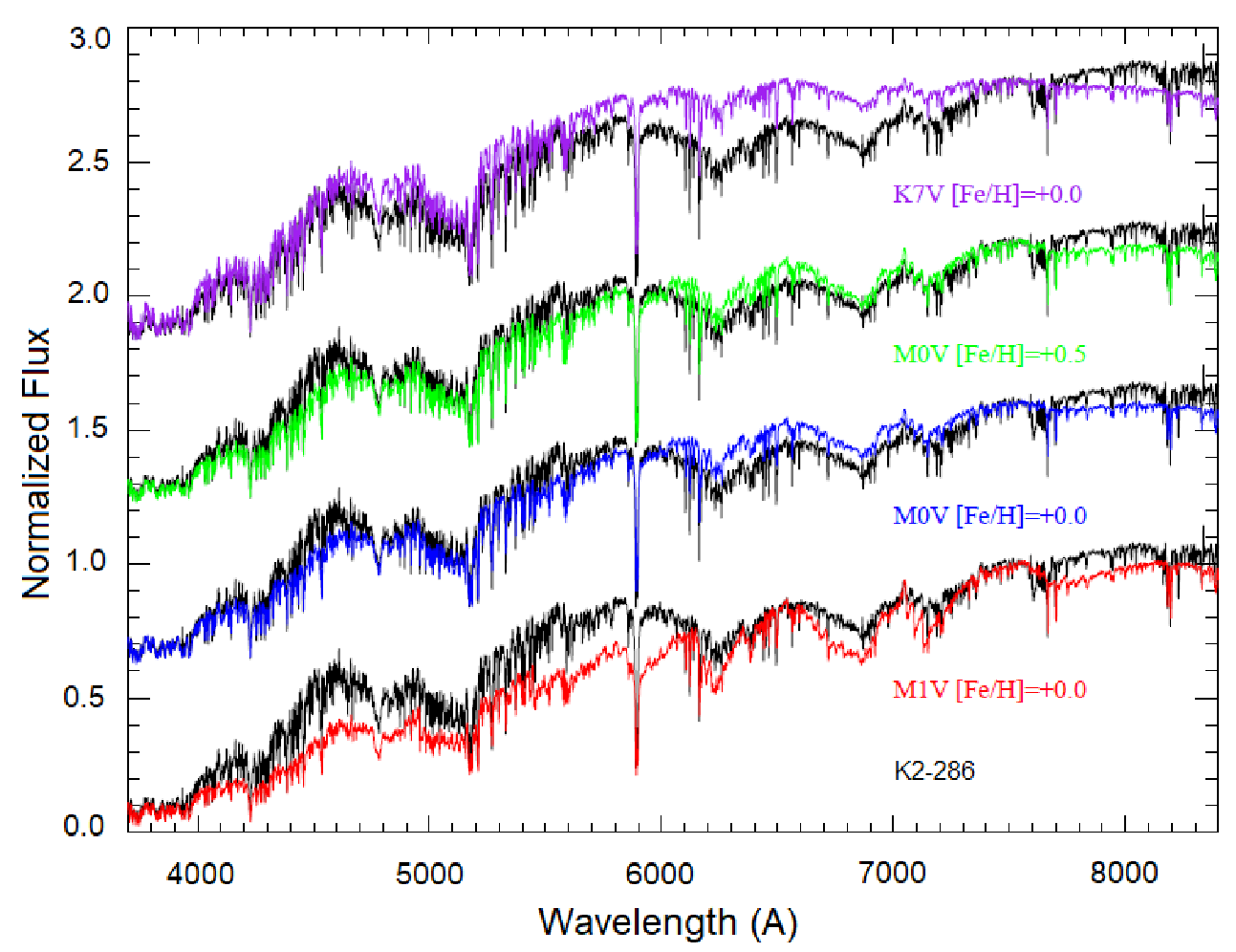}
    \caption{OSIRIS spectrum of K2-286 compared with reference spectra of K7V -- M1V stars with $\mathrm{[Fe/H]} = 0.0$ and with an additional reference spectrum of an M0V star with $\mathrm{[Fe/H]} = +0.5$. All spectra are normalized at $\lambda$ = 7575 {\AA}.}
    \label{fig:epic3_specfit30}
\end{figure*}

The data reduction was performed in a standard way (bias substraction, flat-fielding and wavelength calibration, 
using HgAr$+$NeXe lamps) with the \emph{onedspec} package in IRAF \citep{1993ASPC...52..173T}. The spectra were flux-calibrated using the flux standard GD$-$140 ($\alpha=11:37:22.16$, $\delta=29:48:24.7$ (J2000), $m_{V}=12.45$) provided by GTC team and observed with the same setup and reduced in the same way as our science target. The \emph{standard} IRAF package contains precise spectrophotometric data of GD$-$140 to derive the sensitivity curve of the instrument in that night in each filter. Finally, we corrected the four spectra with the derived sensitivity curves to achieve a reliable flux calibration. The spectra of each filter were corrected for barycentric velocity and merge to produce the full spectrum shown in
Fig.~\ref{fig:epic3_specfit30}.

We compared the resulting spectrum with SDSS/BOSS reference spectra of M-type stars using the HAMMER code \citep{2007AJ....134.2398C} and using the python version PYHAMMER \citep{2017ApJS..230...16K}. The best fit is obtained for an M0V  star with $\mathrm{[Fe/H]} \sim +0.5$, although a chi--squared test finds best fit for $\mathrm{[Fe/H]}\sim0.0$ (see Fig.~\ref{fig:epic3_specfit30}).

To compute the stellar luminosity we used $m_{k}$ (from 2MASS) and the distance from Gaia ($d=76.3\pm0.3$ pc) to obtain $M_{K}=4.90\pm0.02$. Following \cite{2015ApJ...804...64M} we estimated $BC_{K}=2.48\pm0.02$. From $M_{K}$ and $BC_{K}$ we compute $L=0.089\pm0.003$ L$_{\odot}$. We also estimated the luminosity of K2-286 from the tabulated stellar parameters of \cite{2013ApJS..208....9P} and $M_{K}$, obtaining $L=0.091\pm0.004$ L$_{\odot}$. We adopt a mean value of $L=0.090\pm0.005$ L$_{\odot}$ \footnote{Mean uncertainties are computed adding in quadrature the individual uncertainties.}.

We estimated the stellar radius applying the empirical relation $R-M_{k}$ from \cite{2015ApJ...804...64M} with $M_{k}$ computed from $m_{k}$ of 2MASS and distance from Gaia. We obtain $R=0.63\pm0.02$ R$_{\odot}$. We also computed the radius from $M_{k}$ and the \cite{2013ApJS..208....9P} relations, obtaining $R=0.61\pm0.01$ R$_{\odot}$. We adopt a mean value for the stellar radius of $R=0.62\pm0.02$ R$_{\odot}$

We obtained $T_{\rm eff}$ for K2-286 from three different approaches. First, we followed the work of \cite{2015A&A...577A.132M} who calibrated empirical relationships to determine accurate stellar parameters for early-M dwarfs (spectral types M0 -- M4.5) using ratios of pseudo-equivalent widths of spectral features as a temperature diagnostic. With this method we obtain $T_{\rm eff}$ $\sim$ 3820 $\pm$ 50 K for K2-286. Second, following \cite{2017AJ....153..267M} we derived $T_{\rm eff}$ from the Stefan-Boltzmann relation. With the computed $L$ and $R$, we obtain $T_{\rm eff}$ $\sim$ 4015 $\pm$ 85 K. Third, we estimated $T_{\rm eff}$ from optical and infrared photometry following \cite{2008MNRAS.389..585C}, applying the empirical relation between $T_{\rm eff}$ and $(V-J)$, to obtain $T_{\rm eff}$ $\sim$ 3945 $\pm$ 21 K. From these three methods, we adopt a mean value for $T_{\rm eff}$ of 3926$\pm$ 100 K.  \footnote{Gaia DR2 reports $T_{\rm eff}$ $\sim$ 4035 $\pm$ 100 K, while \cite{2013ApJS..208....9P} report a temperature of $\sim$ 3870 K for M0V stars, both consistent with our adopted value for $T_{\rm eff}$ of 3926 $\pm$ 100 K.}

We derived the stellar mass applying the $mass - M_{k}$ relation of \cite{2016AJ....152..141B}, obtaining $M= 0.64\pm0.01$ M$_{\odot}$. We also estimated the mass of K2-286 with the $mass - M_{k}$ relation from \cite{2015ApJ...804...64M}, obtaining $M= 0.65 \pm 0.01$ M$_{\odot}$. Finally, from the tabulated stellar parameters of \cite{2013ApJS..208....9P} we estimate $M= 0.64 \pm 0.01$ M$_{\odot}$. We adopt a mean value for the stellar mass of $M= 0.64 \pm$0.02 M$_{\odot}$.

\begin{figure}
	\includegraphics [width=0.45\textwidth]{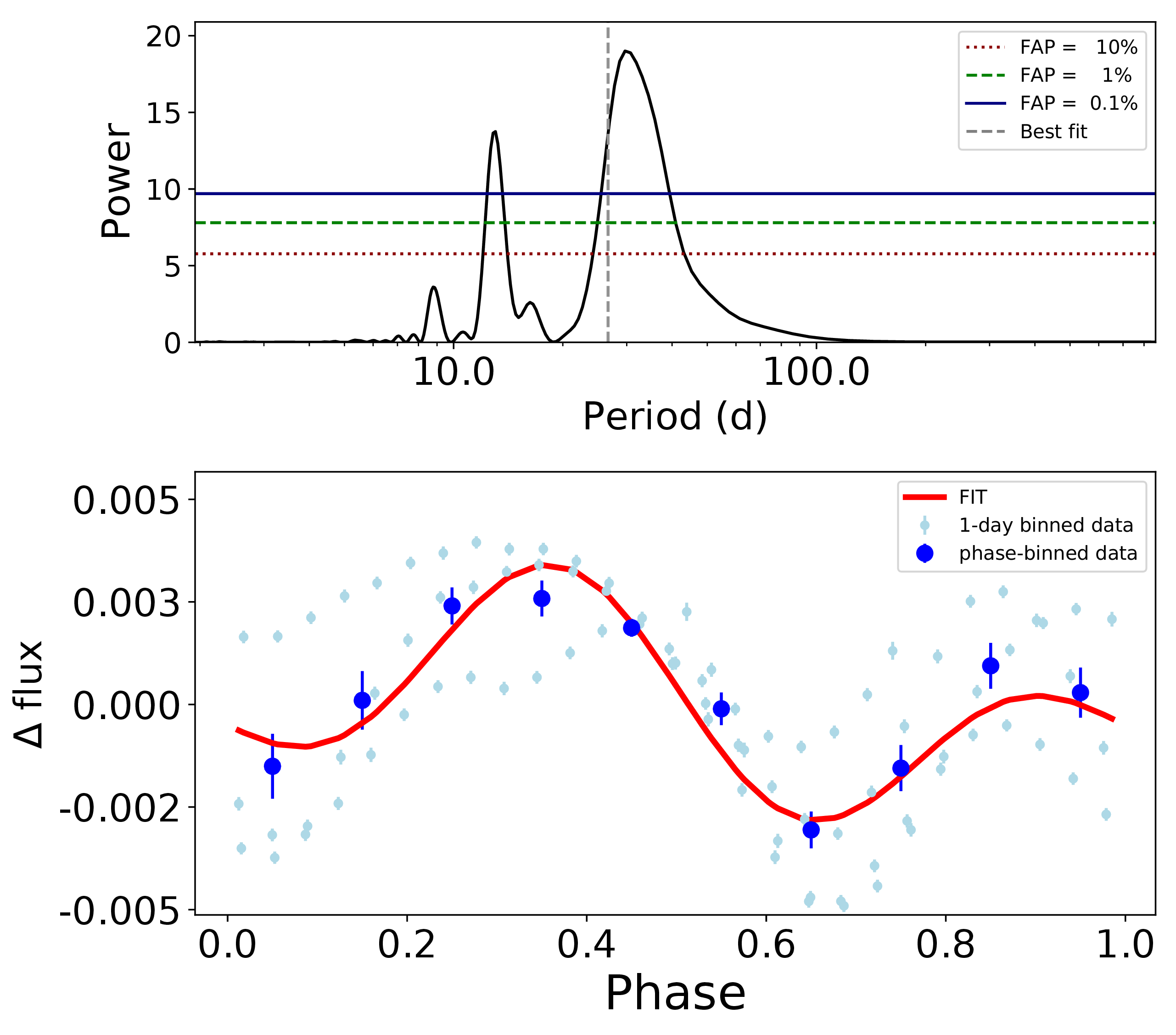}
    \caption{Top panel: GLS periodogram of light curve of K2-286. Strongest signal is present at $26.6\pm0.5$ d. Bottom panel: light curve and fit corresponding to the period of $26.6\pm0.5$ d.}
    \label{fig:rotation}
\end{figure}

The light curve of K2-286 shows modulation associated to rotation with a double dip behavior \citep{2018ApJ...863..190B}. We applied a Generalized Lomb-Scargle \citep{2009A&A...496..577Z} analysis to the light curve of K2-286 to compute the photometric rotation period. We find the strongest signal at $26.6\pm0.5$ d (see Fig.~\ref{fig:rotation}).

K2-286 has not been observed with X-ray telescopes as Chandra or XMM-Newton, but the field of K2-286 was observed in the ROSAT All sky survey for 350 seconds and the star was not detected. With the X-ray flux upper limit of ROSAT at $F_{X}<3.0\cdot10^{-13}$ erg$\cdot$s$^{-1}\cdot$cm$^{-2}$ we estimate $L_{X}<2.1\cdot10^{29}$ erg$\cdot$s$^{-1}$ for K2-286. 

\begin{figure}
	\includegraphics [width=0.5\textwidth]{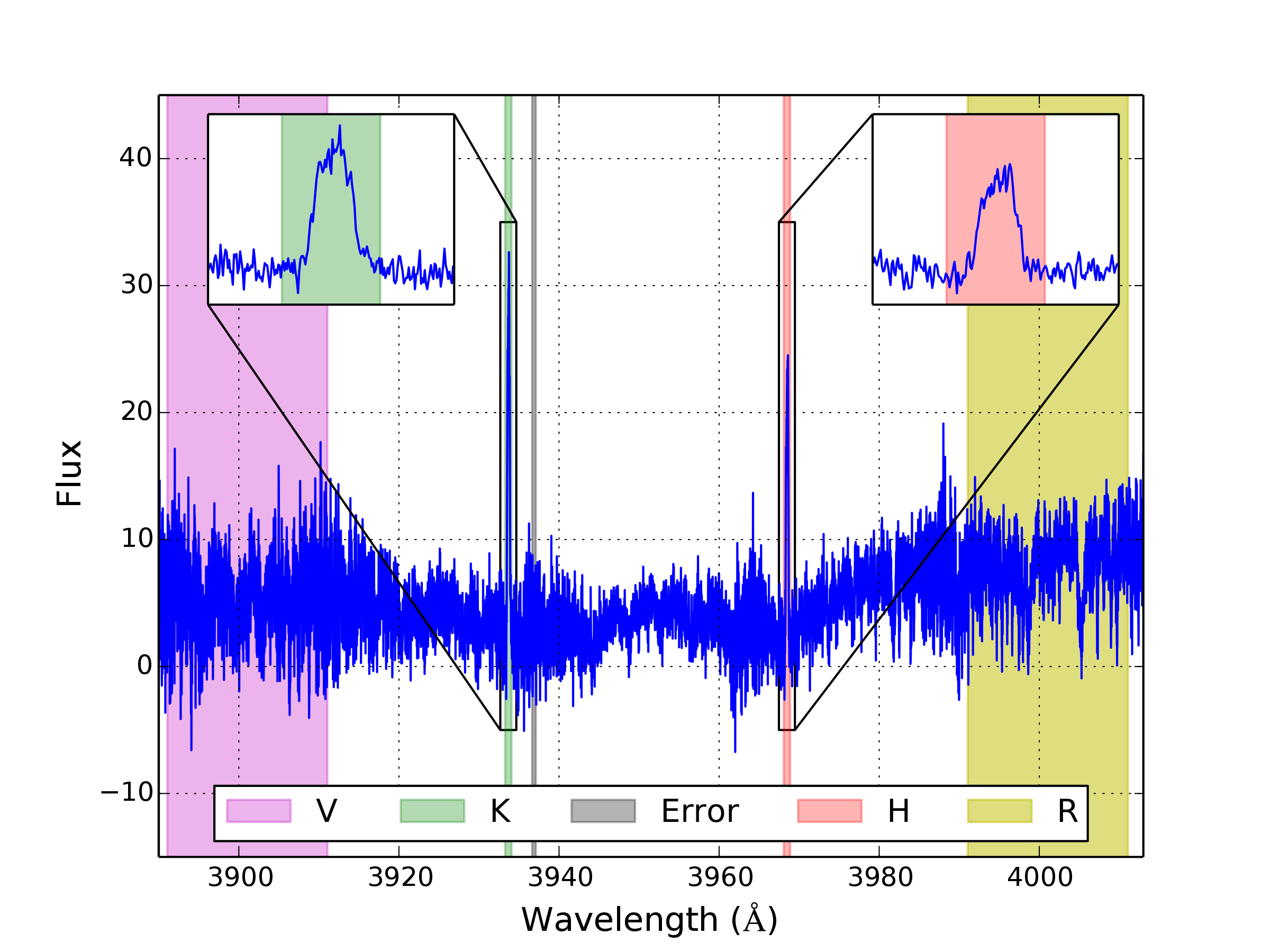}
    \caption{Average one-dimensional spectrum of the six available spectra taken with HARPS-N. The $Ca II H\&K$ filters are marked in pink and green 
respectively including a zoom of those lines; the continuum passbands are marked in violet and yellow and the continuum region used for the 
index error is marked in grey.}
    \label{fig:HARPS}
\end{figure}

Between 2018 July 15 and 2018 August 08 we obtained six spectra, 1800 s of exposure time each, with HARPS-N \citep{2012SPIE.8446E..1VC}, a fibrefed high resolution echelle spectrograph installed at the 3.6 m  Telescopio Nazionale Galileo in the Roque de los Muchachos Observatory (Spain) with  a resolving power of R = 115,000 over a spectral range from  380 to 690 nm.
Fig.~\ref{fig:HARPS} shows the average spectrum obtained by transforming the reduced bidimensional spectra into one dimension, using a custom criteria for the overlapping between the echelle orders instead of using the one-dimensional spectra provided by the HARPS-N pipeline. This is because the overlapping technique used by the pipeline usually includes some of the noisiest parts of the echelle orders, decreasing the signal to noise ratio, which is crucial in the blue orders (where the noise is higher), especially in faint stars.

Following \cite{1984ApJ...279..763N}, we used the average spectrum to measure the CaII H$\&$K index S as S=(H+K)/(R+V), with H, K, R $\&$ V the total flux in each passband, and estimated the mean level of chromospheric activity $\log_{10} (R'_{HK})$ from:

\begin{equation}
R'_{HK}=1.34\cdot 10^{-4}\cdot C_{cf} \cdot S -Rphot
\end{equation}

where $C_{cf}=0.11\pm0.01$ is a conversion factor and $Rphot=40\cdot 10^{-8}\pm4\cdot 10^{-8}$ takes into account the photospheric contribution to this index, both of them dependent on the B and V magnitudes. We obtain $\log_{10} (R'_{HK})=-4.75 \pm 0.06$.

Using the specific relation for M dwarf stars between level of activity and the rotation period found by \cite{2018A&A...612A..89M}: 

\begin{equation}
log_{10}(P_{rot}) = A+B\cdot\log_{10}(R'_{HK})
\end{equation}

with the fitted parameters $A=-2.15\pm0.27$ and $B=-0.731\pm0.055$ , we estimate a rotation period $P_{rot}=21.0\pm3.7$ days for K2-286.

The photometric (from K2) and spectroscopic observations (from HARPS-N) of K2-286 were made in short campaigns, with one year separation. Differential rotation induces variations in the measured photometric rotation period depending on when the data was taken, and the chromospheric estimation of the rotation period depends on the epoch of the stellar magnetic cycle and also on the superficial inhomogeneities responsible for the emission of CaII H$\&$K and their latitudinal location. A combination of these effects explains the difference between the photometric and chromospheric rotation period, for which we finally adopt $P_{rot}=23.8\pm3.7$ days, the mean value of both measurements.

From Gaia DR2 we adopt a distance $d=76.3\pm0.3$ pc, proper motions $\mu_{\alpha}=57.807\pm0.084$ mas$\cdot$yr$^{-1}$ and $\mu_{\delta}=-105.645\pm0.066$ mas$\cdot$yr$^{-1}$, and $v_{r}=-17.41\pm1.17$ km$\cdot$s$^{-1}$. With these parameters we compute velocity components U = 4.9 km$\cdot$s$^{-1}$, V = -11.5 km$\cdot$s$^{-1}$, W = -45.2 km$\cdot$s$^{-1}$.

\cite{2006MNRAS.367.1329R}, using accurate radial velocities and Hipparcos astrometry, deduced a relation for estimating the probability for a star to belong to the galactic thin disk, thick disk or the halo, by means of the velocity components U,V and W. Following their work we conclude that K2-286 is most likely a member of the thin disk with probability $P=0.938$ (the probabilities of being a thick disk or halo member are $P=0.061$ and $P=0.001$ respectively).

All the parameters for K2-286 are listed in Table \ref{tab:stellar_parameters}.

\begin{table}
	  
	\centering
	\caption{Stellar parameters for K2-286}
	\label{tab:stellar_parameters}
    \resizebox{9 cm}{!} 
    {
	\begin{tabular}{lcc}
		       
        Parameter&Value&Source\\
        \hline
        \hline
        V [mag]& $12.763 \pm 0.010$&(1)\\
        R [mag]& $12.196 \pm 0.020$&(1)\\
        I [mag]& $11.628 \pm 0.020$&(1)\\
        J [mag]& $10.127 \pm 0.024$&(2)\\
        H [mag]& $9.518 \pm 0.022$&(2)\\
        K [mag]& $9.317 \pm 0.020$&(2)\\
        $T_{\rm eff}$ [K]& $3926\pm100$ &(4)\\
        $\mathrm{[Fe/H]}$& $0.0 \pm 0.5$&(4)\\              
       	Radius [R$_{\odot}$]&$0.62\pm 0.02$&(4)\\
		Mass [M$_{\odot}$]&$0.64 \pm0.02$&(4)\\
        Luminosity [L$_{\odot}$]&$0.090 \pm 0.05$&(4)\\
        $P_{rot}$ [d]&$23.8\pm3.7$&(4)\\
        $\log g$ [cgs]&$4.7 \pm 0.2$&(4)\\
        Distance [pc] &$76.3 \pm 0.3$&(3)\\
        $\mu_{\alpha}$ [mas$\cdot$yr$^{-1}$]&$57.807\pm0.084$ &(3)\\
        $\mu_{\delta}$ [mas$\cdot$yr$^{-1}$]&$-105.645 \pm 0.066$ &(3)\\
        $V_{\rm r}$ [km$\cdot$s$^{-1}$]&$-17.41 \pm 1.17$ &(3)\\
        U, V, W  [km$\cdot$s$^{-1}$]&4.9, -11.5, -45.2 &(4)\\
       &&\\
        \multicolumn{3}{l}{(1) UCAC4 \citep{2013AJ....145...44Z}.}\\
         \multicolumn{3}{l}{(2) 2MASS \citep{2003tmc..book.....C}.}\\
         \multicolumn{3}{l}{(3) Gaia DR2 \citep{2018arXiv180409365G}.}\\
         \multicolumn{3}{l}{(4) This work.}\\
	\end{tabular}
	}
\end{table}

\subsection{K2 photometric data}

\begin{figure*}
	\includegraphics [width=0.9\textwidth]{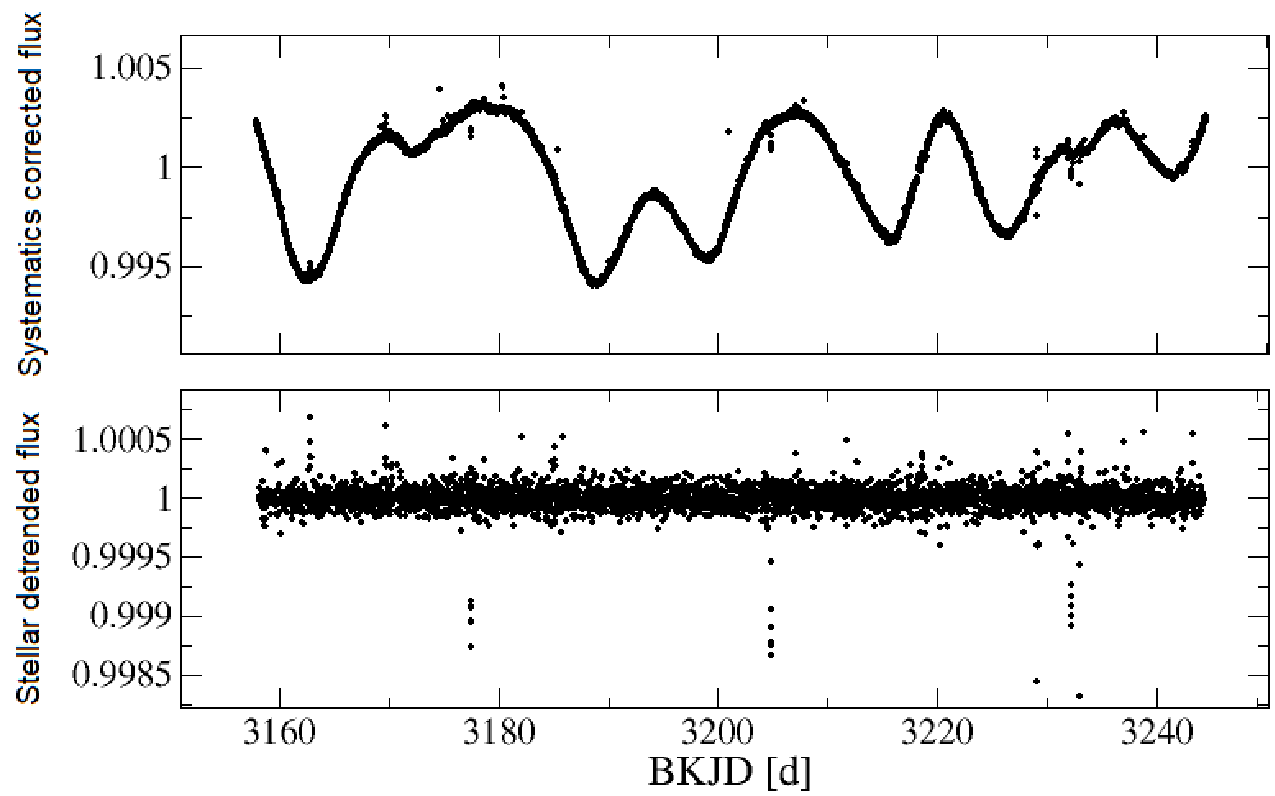}
    \caption{K2 systematics corrected (top) and stellar detrended (bottom) light curves for K2-286.}
    \label{fig:lightcurve}
\end{figure*}

Following the work of \cite{2014PASP..126..948V} we analyzed the K2 corrected photometry of K2-286. The stellar variability was detrended with a moving median filter before searching for periodic signals using the Box Least Squares (BLS) method \citep{2002A&A...391..369K} on attained data. The analysis shows a transit signal with period of $27.359 \pm 0.005$ days (Fig.~\ref{fig:lightcurve}). Three transit events corresponding to this signal are present in the light curve.

\begin{figure}
	\includegraphics [width=0.5\textwidth]{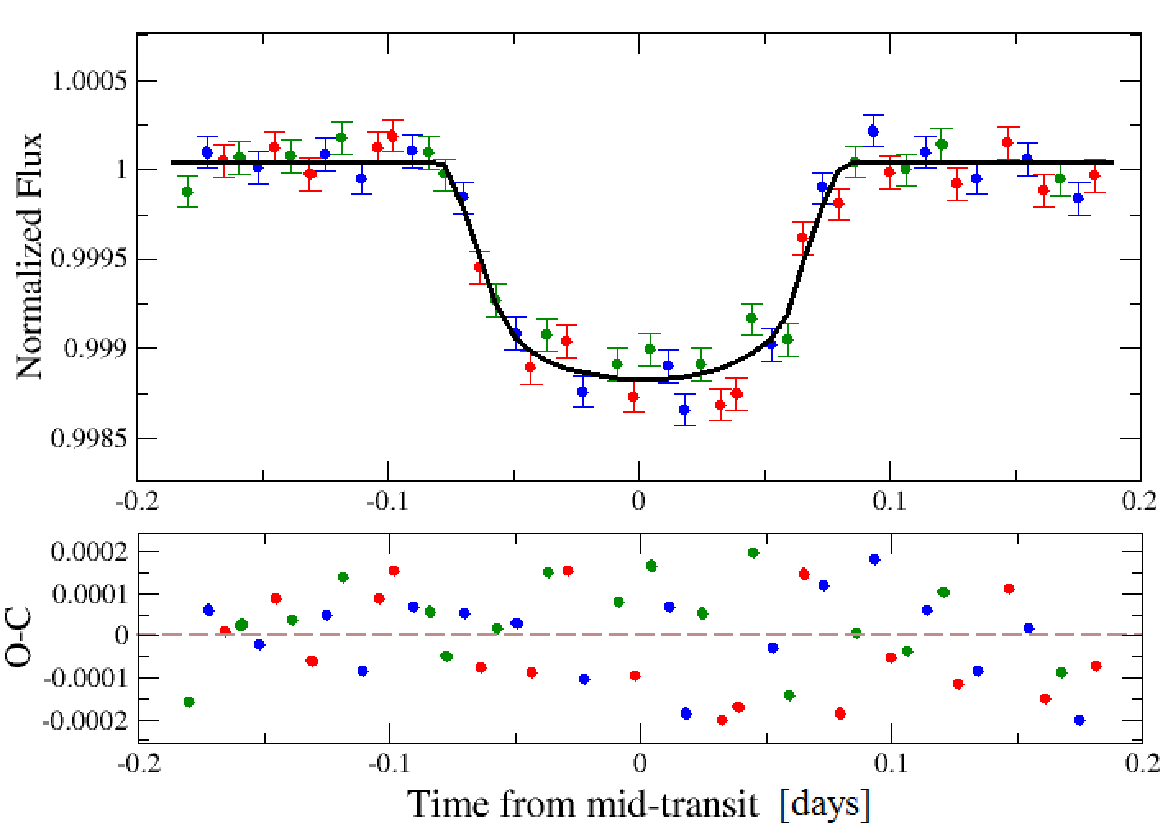}
    \caption{Top: Phase-folded light curve of planet K2-286b (red, blue and green dots correspond to first, second and third transit). Solid curve represents best model fit obtained by MCMC. Bottom: residuals of the fit.}
    \label{fig:transitoplanetab}
\end{figure}

To estimate the parameters for K2-286b we performed MCMC analysis on the phase-folded transit (Fig.~\ref{fig:transitoplanetab}), using the PYANETI package \citep{2017ascl.soft07003B} to fit models from \cite{2002ApJ...580L.171M} and following  \cite{2010MNRAS.408.1758K} to treat the 29.5-minute long cadence of K2. For the calculations we set the values of $T_{\rm eff}$, $R$ and $M$ for K2-286 listed in Table~\ref{tab:stellar_parameters}. We worked with the assumption of eccentricity $e=0$ as the light curve alone does not constrain properly the eccentricity. We also assumed Gaussian priors for the linear and quadratic limb darkening coefficients $u1=0.52$, $u2=0.22$ \citep{2010A&A...510A..21S} with 0.1 standard deviation.

The analysis resulted in an estimated radius of $2.1\pm0.2$ R$_{\oplus}$, semi-major axis  $0.1768^{+0.0175}_{-0.0205}$ AU and orbital period $27.359\pm0.005$ days. Table~\ref{tab:planetparams} summarizes all the parameters obtained for K2-286b.

\subsection{False positives analysis}

\begin{figure}
	\includegraphics [width=0.45\textwidth]{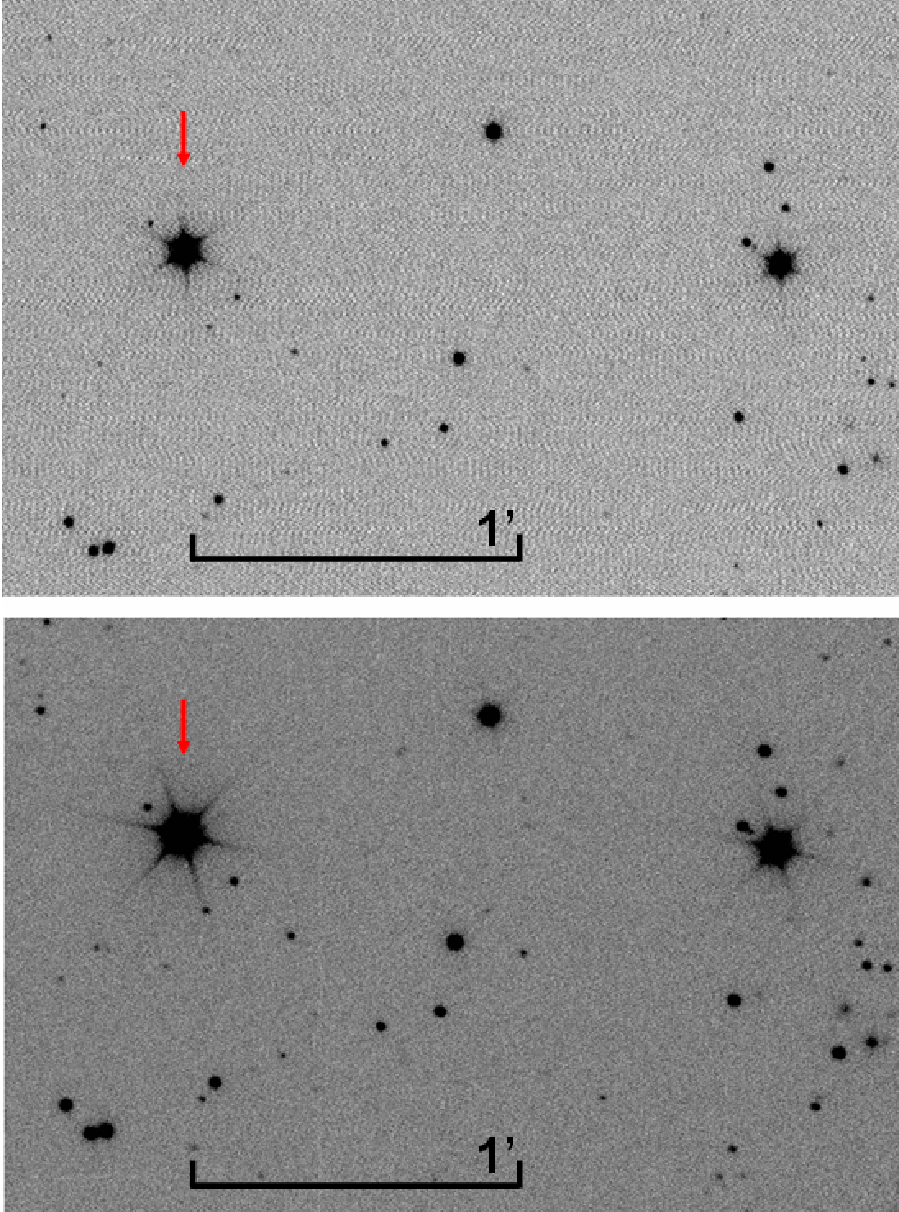}
    \caption{OSIRIS/GTC images of K2-286 field from 11 x 0.1 sec (top) and 11 x 0.5 sec (bottom) series in the $i$-band Sloan filter. Arrows point K2-286. The closest three stars clearly seen, located at 8.4, 13.4 and 15.2 arc seconds, have magnitude differences with respect to the main target of roughly 7.8, 7.2 and 7.7 mag.}
    \label{fig:OSIRISAR}
\end{figure}

\begin{figure}
	\includegraphics [width=0.45\textwidth]{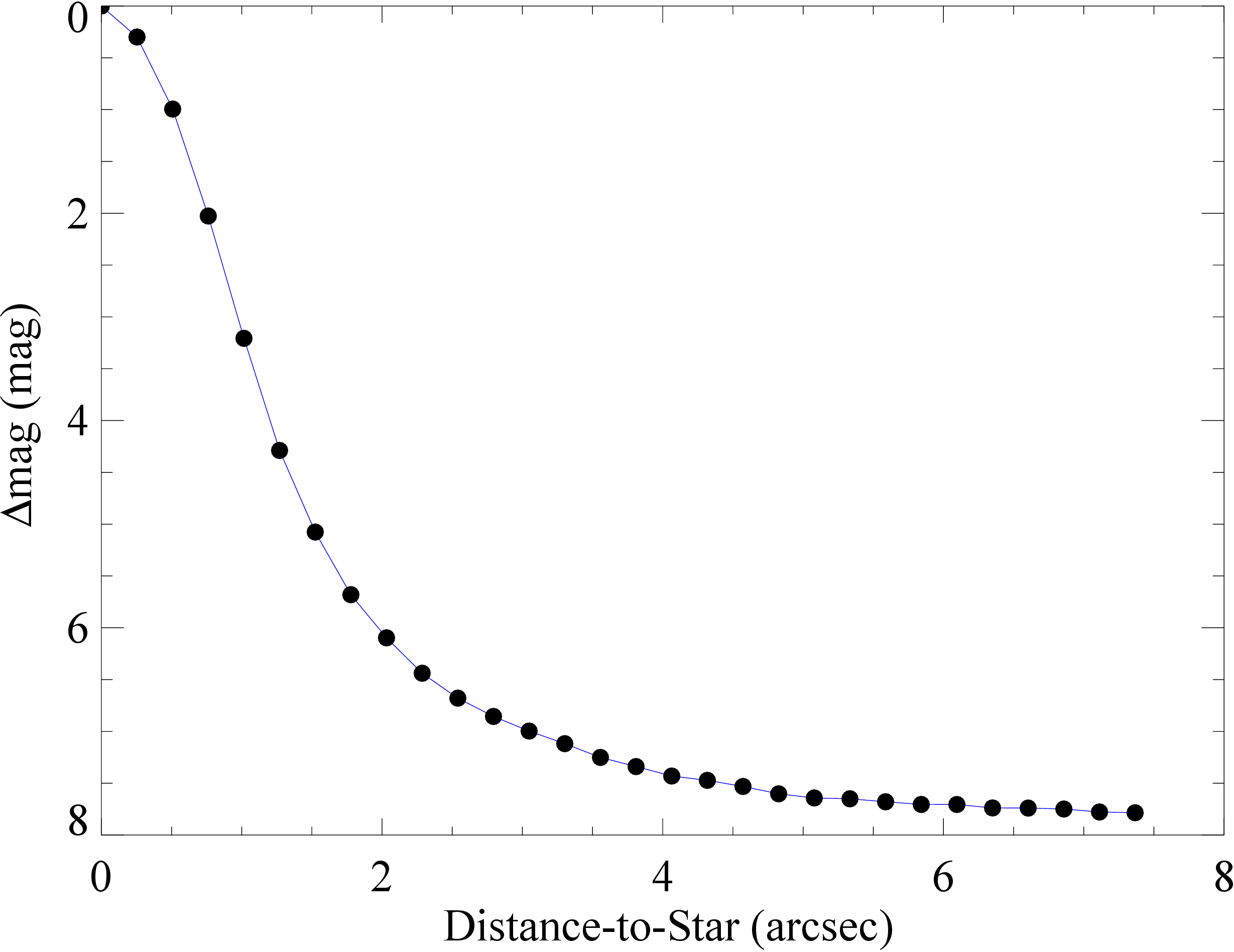}
    \caption{Contrast curve for K2-286 from the OSIRIS/GTC image.}
    \label{fig:contrastcurve}
\end{figure}

We acquired images of K2-286 with the OSIRIS camera-spectrograph on June 19 2018. Data were collected under photometric conditions, with an average seeing of 0.7 arc seconds. Two series of 11 x 0.5 and 11 x 0.1 sec in Sloan $i$ filter were obtained for broadband imaging. Bias correction, flat fielding and bad pixel masking were done using standard procedures. The images were finally aligned. Analysis of the final image excludes companions at 1.5 arc seconds with $\delta$mag < 5.0 and at 3 arc seconds with $\delta$mag < 7 (see Figs.~\ref{fig:OSIRISAR} \&~\ref{fig:contrastcurve}).

\begin{figure}
	\includegraphics [width=0.45\textwidth]{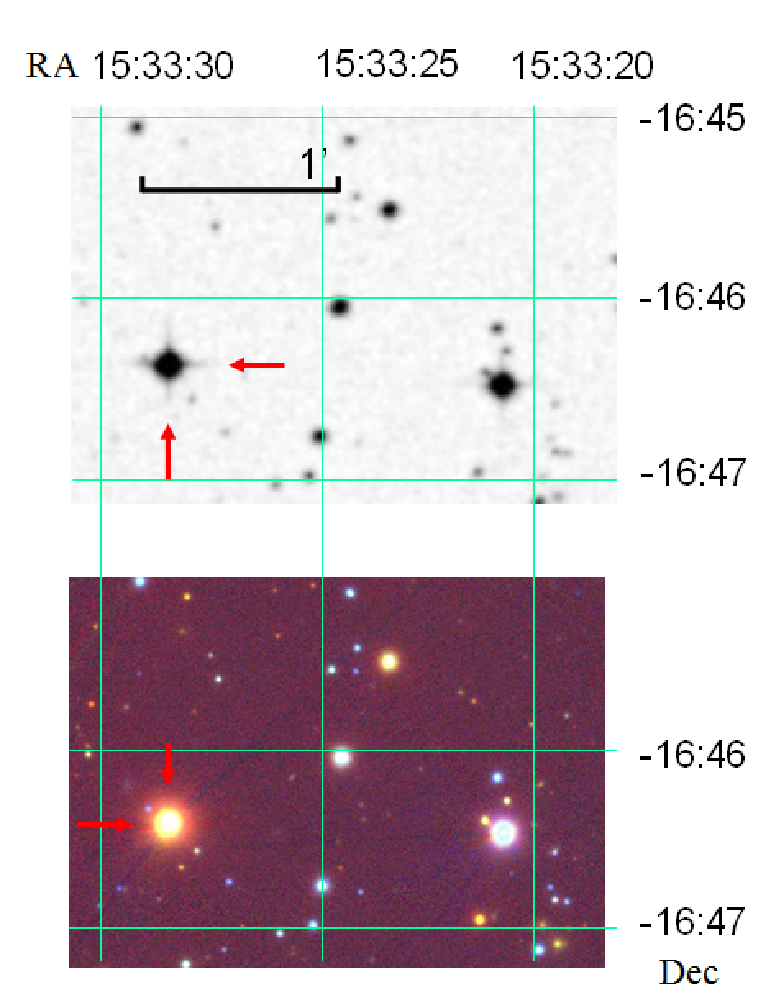}
    \caption{POSS-I image (1953, top) and $y/i/g$ stack from PanSTARRS-1 (2014, bottom). Arrows point K2-286.}
    \label{fig:PANSTARRS_POSSI}
\end{figure}

Inspection of images of K2-286 from POSS-I \citep[][year 1953]{1963bad..book..481M} 2MASS \citep[][year 1998]{2003tmc..book.....C} and PanSTARRS-1 \citep[][year 2014]{2016arXiv161205560C}
do not show background sources at the current star position (see Fig.~\ref{fig:PANSTARRS_POSSI}).

We have also used the package VESPA \citep{2012ApJ...761....6M,2015ascl.soft03011M} to perform a statistical validation of K2-286b. Using models of stellar populations in the Galaxy, the package computes the probabilities of planetary and non-planetary scenarios, taking into account eclipsing binaries, background eclipsing binaries and hierarchical triple systems scenarios. Running VESPA, we obtain FPP < $10^{-4}$ for K2-286b.

Non-detection of companions or background sources in OSIRIS/GTC, POSSI, 2MASS and PanSTARRS-1 images, and statistical analysis with VESPA, strongly support the planetary origin of the signals present in the light curve of K2-286.

\begin{table*}
	  
	\centering
	\caption{Parameters for planet K2-286b}
	\label{tab:planetparams}
    \resizebox{12 cm}{!} 
    {
	\begin{tabular}{lc}

        Planet parameter&Value\\
        \hline
        \hline
        Orbital period (P) [d]&$27.359\pm0.005$\\
        \hline
        Semi-major axis (a) [AU]&$0.1768^{+0.0175}_{-0.0205}$\\
        \hline
        Radius ($R_{p}$) [R$_{\oplus}$]&$2.1\pm0.2$\\
        \hline
        Mass ($M_{p}$) [M$_{\oplus}$] (1)&$6.8\pm4.3$\\
        \hline
        Equilibrium Temperature ($T_{eq}$) [K]& $347^{+21}_{-11}$\\
        &\\
        Transit parameter&Value\\
        \hline
        \hline
        Epoch (BKJD) [days]&$3177.417 \pm 0.002$\\
        \hline
        Radius of planet in stellar radii ($R_{p}$/$R_{*}$)&$0.0315^{+0.0008}_{-0.0005}$\\
        \hline
        Semi major axis in stellar radii  ($a/R_{*}$)&$63^{+4}_{-7}$\\
        \hline
       	Linear limb-darkening coeff ($u_{1}$)&$0.52\pm0.01$\\
        \hline
		Quadratic limb-darkening coeff ($u_{2}$)&$0.22\pm0.01$\\
        \hline
        Inclination (i) [deg]&$89.67^{+0.21}_{-0.25}$\\
        \hline
        Impact Parameter (b)&$0.35^{+0.21}_{-0.23}$\\
        \hline
        Transit depth ($\delta$)&$0.00112^{+0.00026}_{-0.00016}$\\
        \hline
        Total duration ($T_{14}$) [d]&$0.125^{+0.031}_{-0.017}$ \\
        
        &\\
        &\\
\multicolumn{2}{l}{\small (1): From \cite{2016ApJ...825...19W} and \cite{2017ApJ...834...17C} mass-radius relations.}\\          
	\end{tabular}
}
\end{table*}

\subsection{Habitability}

Assuming a Bond albedo $A=0.3$ and an Earth-like greenhouse effect, we estimate a surface temperature of $376^{+26}_{-19}$ K for K2-286b. However, different models predict different distances for the inner and outer edges of the habitable zone (HZ) of a star.

In this work we compare the conservative model from \cite{2013ApJ...765..131K}, that explores the HZ for planets with atmospheres similar to the Earth's atmosphere, with the optimistic model from \cite{2013ApJ...778..109Z}, that studies the HZ for planets with atmospheres not necessarily like the Earth's atmosphere, but any combination able to support liquid water. In the inner edge of this optimistic model, a desert planet with small or moderate reserves of water at the poles could generate high albedo, very little greenhouse effect, and very little loss of water.

For the conservative model of \cite{2013ApJ...765..131K} we find a distance $d=0.241$ AU for the inner edge of the habitable zone of K2-286. For the optimistic model of \cite{2013ApJ...778..109Z} and assuming albedo $A=0.2$, we estimate the inner edge of the HZ at a distance $d=0.179$ AU. If we assume albedo $A=0.8$ we obtain $d=0.121$ AU. 

\begin{figure*}
	\includegraphics [width=1.0\textwidth]{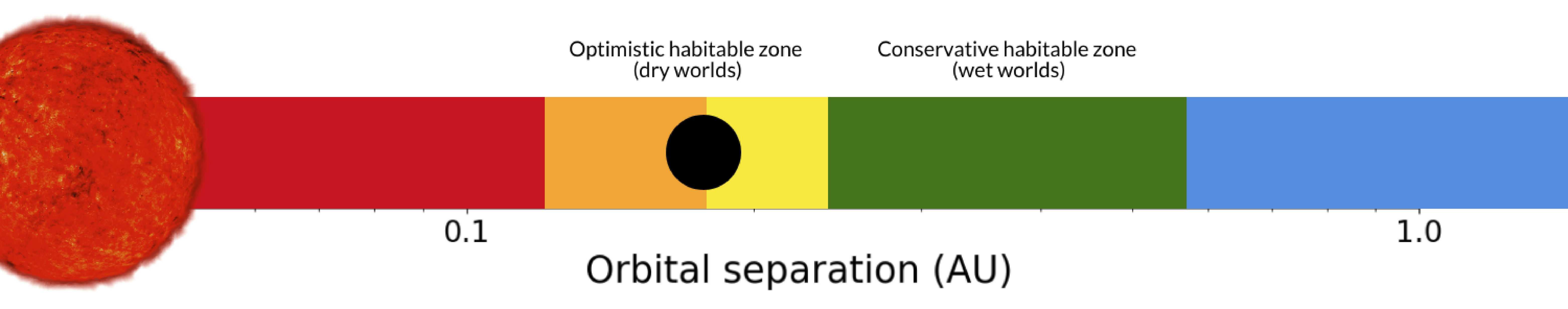}
    \caption{Orange: inner edge of the optimistic habitable zone of K2-286 with albedo A=0.8. Yellow: same with A=0.2. Green: habitable zone limits for the conservative model. Black filled circle represents the location of K2-286b.}
    \label{fig:HZ}
\end{figure*}

With an orbital radius of $0.1768^{+0.0175}_{-0.0205}$ AU, K2-286b would be outside the habitable zone in the conservative model, but within the HZ in the optimistic model (see Fig~\ref{fig:HZ}). These estimations show that K2-286b might be able to support liquid water with the adequate atmospheric conditions.

As stated in section~\ref{sec:stechar}, K2-286 has not been detected as an X-ray source by ROSAT, setting an upper limit for its X-ray luminosity of $L_{X}<2.1\cdot10^{29}$ erg$\cdot$s$^{-1}$. The estimated $\log_{10} (R'_{HK})=-4.75 \pm 0.06$ does not indicate strong activity. Also, visual inspection of the light curve of K2-286 during the 15 campaign observing window of K2, doesn't show strong flare activity. With a host star of spectral type M0V, K2-286b may experience a more benign environment than other planets orbiting M dwarf stars of later types, as Proxima Centauri (M5.5V) and TRAPPIST-1 (M8.0V) systems.

\subsection{Mass estimation}

\cite{2017AJ....154..109F} analyzed the size distribution of a sample of 2025 Kepler planets with precise measured radius, and detected a gap in the radius distribution of close-in planets (orbital period $< 100$ days) at 1.5--2.0 R$_{\oplus}$. \cite{2018AJ....156..264F} confirmed this result using a catalog of $\sim$1000 planets with precise properties from Gaia parallaxes. This gap may be related to photoevaporation, gas poor formation or impact erosion \citep{2017AJ....154..109F}.
Our estimated radius for K2-286b is $R=2.1\pm0.2$ R$_{\oplus}$, in the upper limit of the gap. A precise measurement for the mass of K2-286b is a key point to understand its composition, internal structure, and to test the planetary formation and evolution models.

\begin{figure*}
	\includegraphics [width=1.0\textwidth]{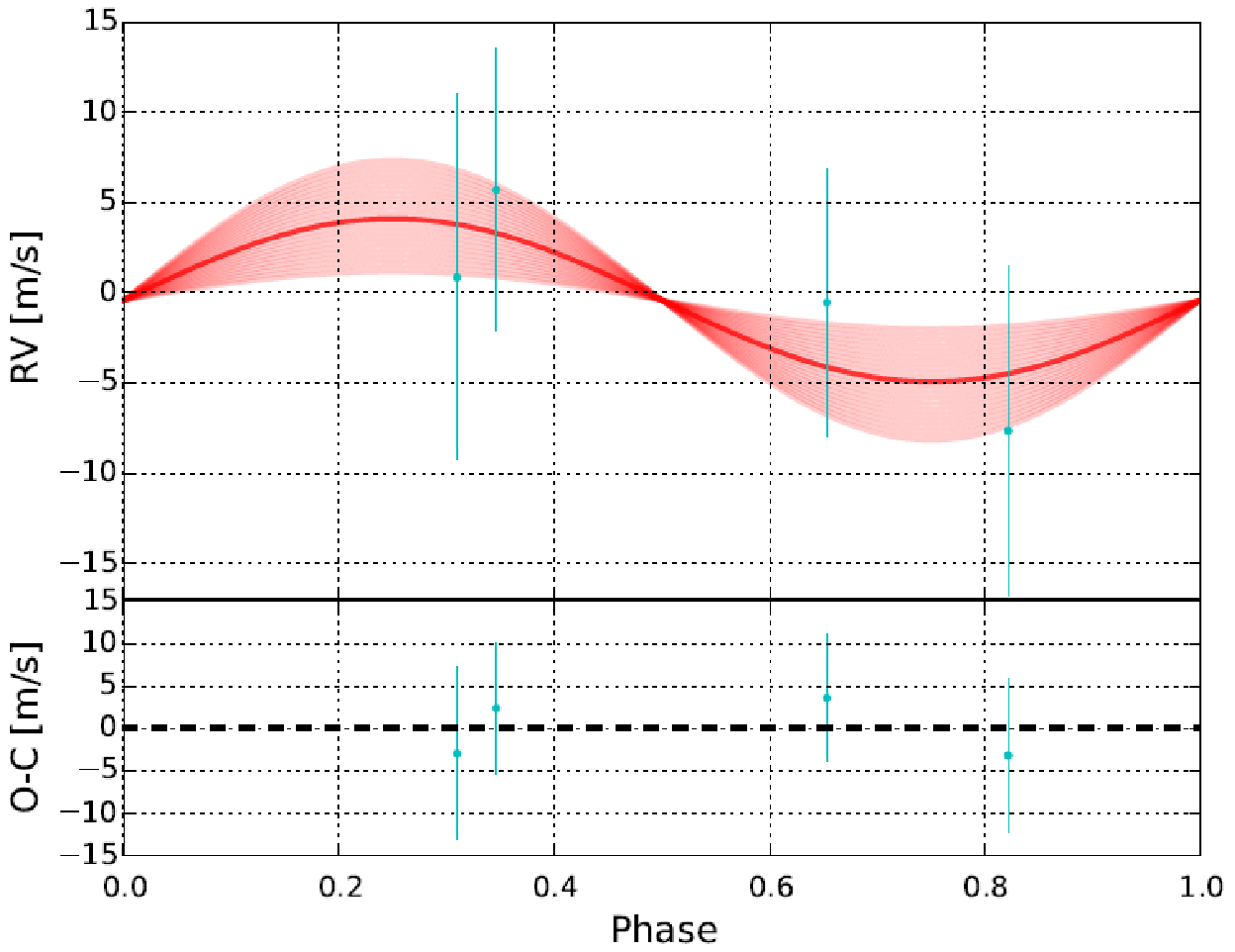}
    \caption{Top panel: HARPS-N $RV$ measurements against orbital phase with the period of 27.359 days and mid-transit time of Tc = 3177.417 (BKJD). The 0.5 phase value corresponds to the epoch of mid-transit. The value at $\sim0.65$ phase is the weighted mean of three spectra acquired consecutively the same night. Overlaid is the best sinusoidal fit with a $RV$ semi-amplitude of $4.52^{+3.29}_{-2.94}$ m$\cdot$s$^{-1}$ and a distribution of other fits for various semi-amplitudes according to the MCMC results (see Fig.~\ref{fig:MCMC}). Bottom panel: residuals from the fit.}
    \label{fig:Vr}
\end{figure*}

In Fig.~\ref{fig:Vr} we plot the $RV$ measurements obtained from the HARPS-N observations against orbital phase with the period of 27.359 days. To obtain the $RV$ values we first fit the 
cross-correlation function given by the HARPS-N pipeline, applying weights to each echelle order and using a Gaussian model \citep{1996A&AS..119..373B}. Then we average the values per night and subtract the weighted average $RV$ to obtain the relative values.
The relative $RV$ values are fit by a sinusoid with the period and phase as fixed parameters. The only free parameter is the $RV$ semi-amplitude, 
which is obtained by an MCMC simulation using the python emcee package \citep{2013PASP..125..306F}. The results of this simulation are shown 
in Fig.~\ref{fig:Vr}. The figure presents various sinusoidal fits with semi-amplitude values that cover the error range given by the MCMC. The best sinusoidal fit is obtained for a semi-amplitude of $4.52^{+3.29}_{-2.94}$ m$\cdot$s$^{-1}$.

Following \cite{2018A&A...612A..89M} we estimate that the semi-amplitude of the period signal induced by the rotation/activity  of the star is $4.4$ m$\cdot$s$^{-1}$. In order to impose an upper limit on the planet signal contribution to the measured semi-amplitude, we quadratically subtract the expected magnetic signal ($4.4$ m$\cdot$s$^{-1}$) from the maximum semi-amplitude compatible with the MCMC determination ($7.8$ m$\cdot$s$^{-1}$). In spite of the very similar rotational period of the star and orbital period of the planet, we do not know if the induced signals are in phase, so signals were quadratically subtracted. We obtain that the maximum amplitude of the planet signal is $6.4$ m$\cdot$s$^{-1}$, which leads to a maximum mass of 22 M$_{\oplus}$ and excludes that K2-286b is a gas giant. More $RV$ data would certainly be needed to obtain a mass determination.

\begin{figure}
	\includegraphics [width=0.5\textwidth]{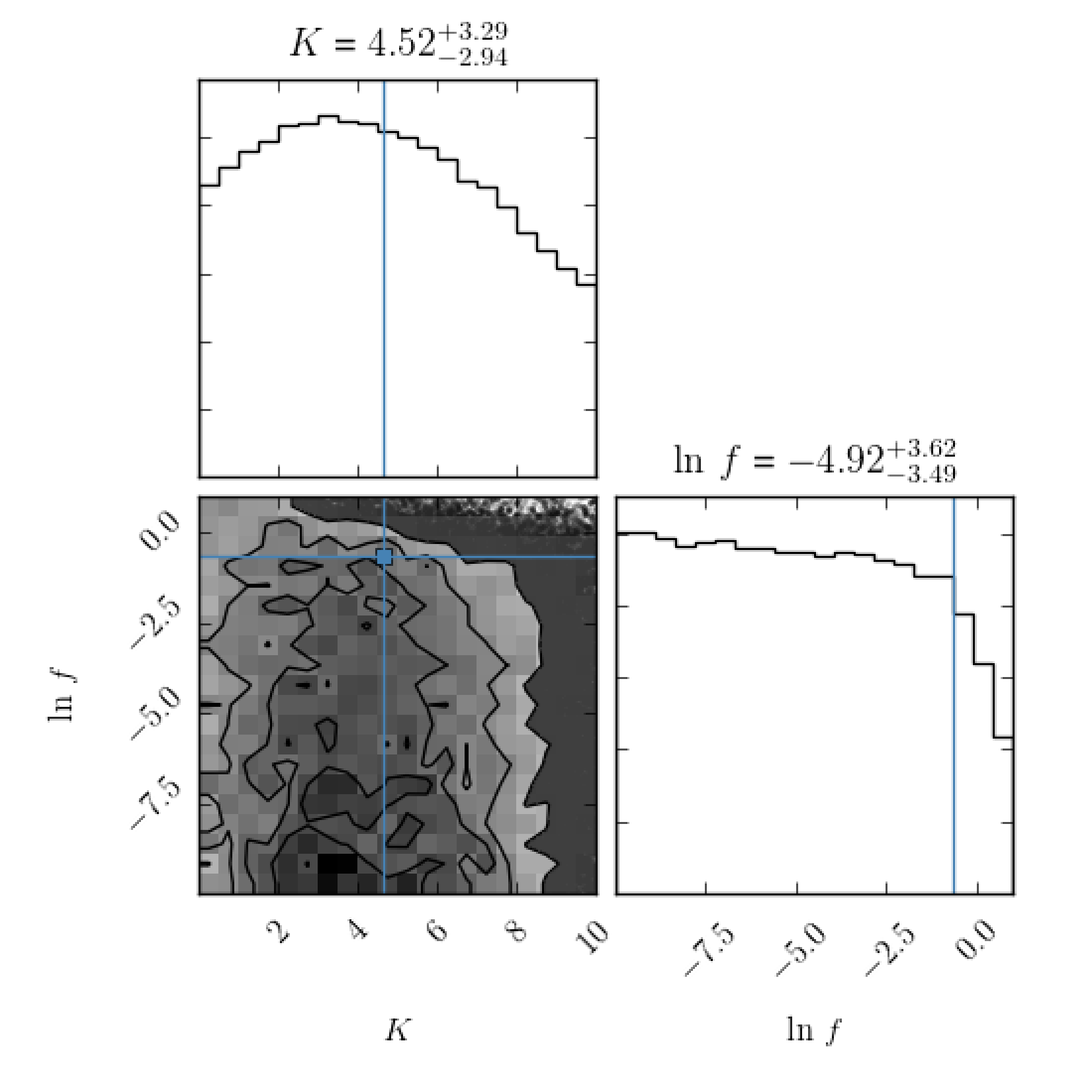}
    \caption{Results of the MCMC fit of the $RV$ values, with $K$ the radial velocity semi-amplitude, and the likelihood parameter $\ln f$ (see Fig.~\ref{fig:Vr}).}
    \label{fig:MCMC}
\end{figure}

Since the previous estimation provides only an upper limit, we estimated the mass of K2-286b applying the following alternative methods:

First, using the probabilistic mass-radius relation from \cite{2016ApJ...825...19W}, which quantifies the intrinsic dispersion and the uncertainties on the relation parameters. Assuming that the relation can be described as a power law with a dispersion that is constant and normally distributed, they found that:

\begin{equation}
\frac{M_{p}}{M_{\oplus}} \sim N \Big(\mu = C\cdot\Big(\frac{R_{p}}{R_{\oplus}}\Big)^{\gamma}, \sigma = \sigma_{M}\Big)
\end{equation}

Adopting $C=2.7$, $\gamma=1.3$ and $\sigma_{M}=1.9$ from the \cite{2016ApJ...825...19W} baseline dataset best fit, we find $M=7.1\pm1.9$ M$_{\oplus}$ for K2-286b.

Second, applying the probabilistic mass -- radius relation implemented in the package FORECASTER \citep{2017ApJ...834...17C}, we obtained $M=6.4\pm3.9$  M$_{\oplus}$.

From these two methods we estimate a mean value for the mass of K2-286b of $6.8\pm$ 4.3 M$_{\oplus}$, within the limits established by our measure of the mass discussed previously.

\section{Discussion}

\begin{figure}
	\includegraphics [width=0.45\textwidth]{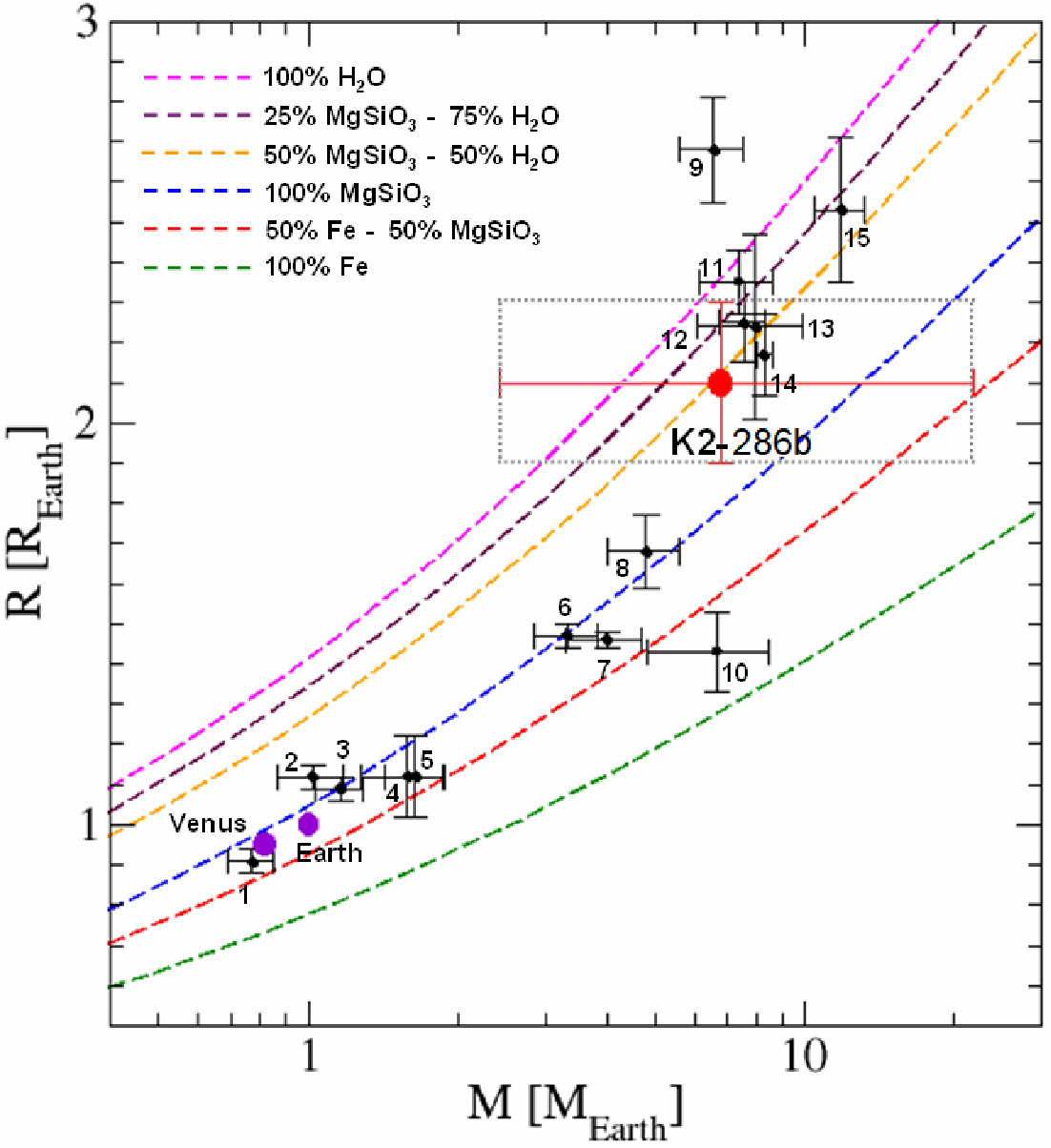}
    \caption{Position of K2-286b (red dot) in the planetary mass - radius space \citep{2013PASP..125..227Z}, 
    adopting the estimated mass from discussed mass-radius relations. Dotted area shows the conservative upper mass limit from the HARPS-N measurements, lower mass limit from probabilistic empirical relations, and radius limits from MCMC analysis of the phase-folded transit. Representative transiting exoplanets with well measured radius and mass are also plotted. The numbers indicate planets (1,2,3) Trappist-1e,b,c \citep{2018A&A...613A..68G}, (4) Kepler-78b \citep{2013Natur.503..377P}, (5) GJ-1132b \citep{2015Natur.527..204B}, (6) Kepler-10b \citep{2014ApJ...789..154D}, (7) Kepler-93b \citep{2015ApJ...800..135D}, (8) Corot-7b \citep{2014A&A...569A..74B}, (9) GJ-1214b \citep{2009Natur.462..891C}, (10) LHS 1140b \citep{2017Natur.544..333D}, (11) Kepler-10c \citep{2017MNRAS.471L.125R,2018ApJ...866...99B}, (12) HD 97658b \citep{2014ApJ...786....2V}, (13) K2-18b \citep{2017A&A...608A..35C}, (14) 55 Cnc e \citep{2012ApJ...759...19E}, (15) HIP 116454b \citep{2015ApJ...800...59V}.}
    \label{fig:RvsM}
\end{figure}

Planets with $R<1.6$ R$_{\oplus}$ have densities consistent with purely rocky composition \citep{2014ApJ...783L...6W,2015ApJ...801...41R}, while planets with $R > 1.6$ R$_{\oplus}$ have densities that suggest a rocky core surrounded by a gaseous envelope \citep{2014ApJ...783L...6W,2015ApJ...801...41R}. Fig.~\ref{fig:RvsM} shows the possible location  of K2-286b in the planetary mass -- radius space \citep{2013PASP..125..227Z}. A sample of representative transiting exoplanets with well measured radii and masses are also plotted. The uncertainties in mass and radius allow a wide range for the composition of K2-286b, excluding only a $100\%$ Fe planet.

\begin{figure}
	\includegraphics [width=0.45\textwidth]{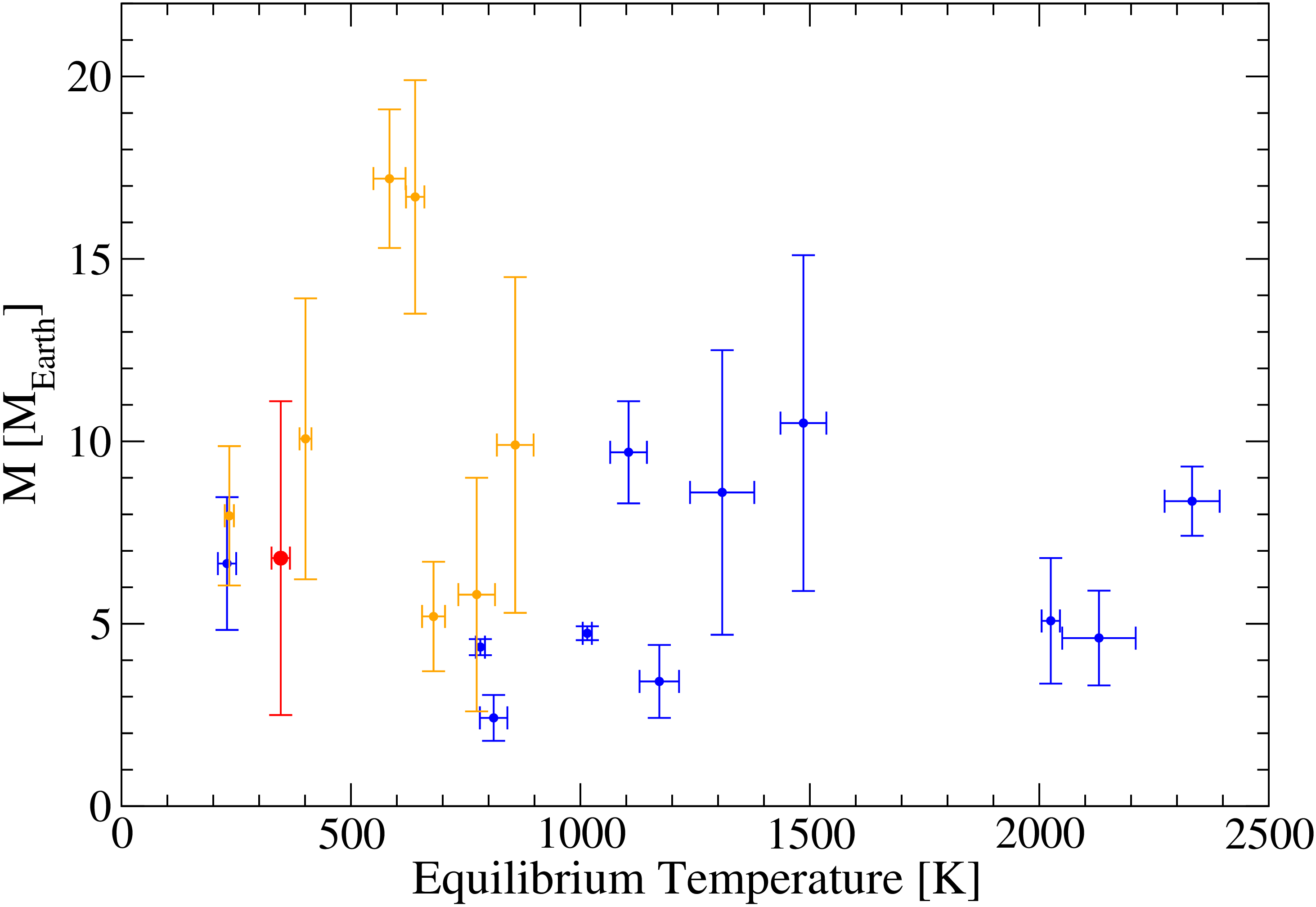}
    \caption{Mass versus $T_{eq}$ for transiting exoplanets with $R<3$ R$_{\oplus}$ and well measured mass. Blue dots: planets with 1 R$_{\oplus}<R<$ 2 R$_{\oplus}$, orange dots: 2 R$_{\oplus}<R<$ 3 R$_{\oplus}$. Red dot: K2-286b. Estimated mass from discussed mass-radius relations is adopted for K2-286b.}
    \label{fig:Teqvsmass}
\end{figure}

In Fig.~\ref{fig:Teqvsmass} we plot $T_{eq}$ against mass for a sample of transiting exoplanets with $R<3$ R$_{\oplus}$ and with well measured masses obtained from radial velocity follow--up (the sample was obtained from the Exoplanet Archive \footnote{https://exoplanetarchive.ipac.caltech.edu/} to date 2018 September 9). K2-286b is plotted adopting the estimated mass from the mass-radius relations discussed previously. The plot shows that K2-286b is cooler than most of them, so a precise measurement of the mass of K2-286b would be of special interest for a better understanding of the mass - radius relation over a wider range of temperatures.

With the estimated mass for K2-286b of $6.8\pm$ 4.3 M$_{\oplus}$ and the assumption of $M_{p} \ll M_{*}$, circular orbit and $\sin i\sim$1, we estimate the induced semi-amplitude in stellar velocity variations at $1.9^{+1.3}_{-1.2}$ m$\cdot$s$^{-1}$. 

The measured $RV$ semi-amplitude of $4.52^{+3.29}_{-2.94}$ m$\cdot$s$^{-1}$ from HARPS-N and the estimated contribution of $4.4\pm0.5$ m$\cdot$s$^{-1}$ due to the stellar activity are consistent, within the uncertainties, with the estimation of $1.9^{+1.3}_{-1.2}$ m$\cdot$s$^{-1}$ for K2-286b, and therefore with the estimated mass of $6.8\pm$ 4.3 M$_{\oplus}$.

The estimated induced semi-amplitude in stellar velocity variations for K2-286b is well-suited  for radial velocity monitoring with ultra-stable spectrographs such as ESPRESSO at the VLT, which is expected to reach $\sim 0.5$ m$\cdot$s$^{-1}$ $RV$ precision for a $m_{V}\sim 13.0$ star in 60 minutes in the single HR mode ($R \sim 134000$), or EXPRES for the Discovery Channel Telescope \citep{2016SPIE.9908E..6TJ}.

\begin{figure}
	\includegraphics [width=0.45\textwidth]{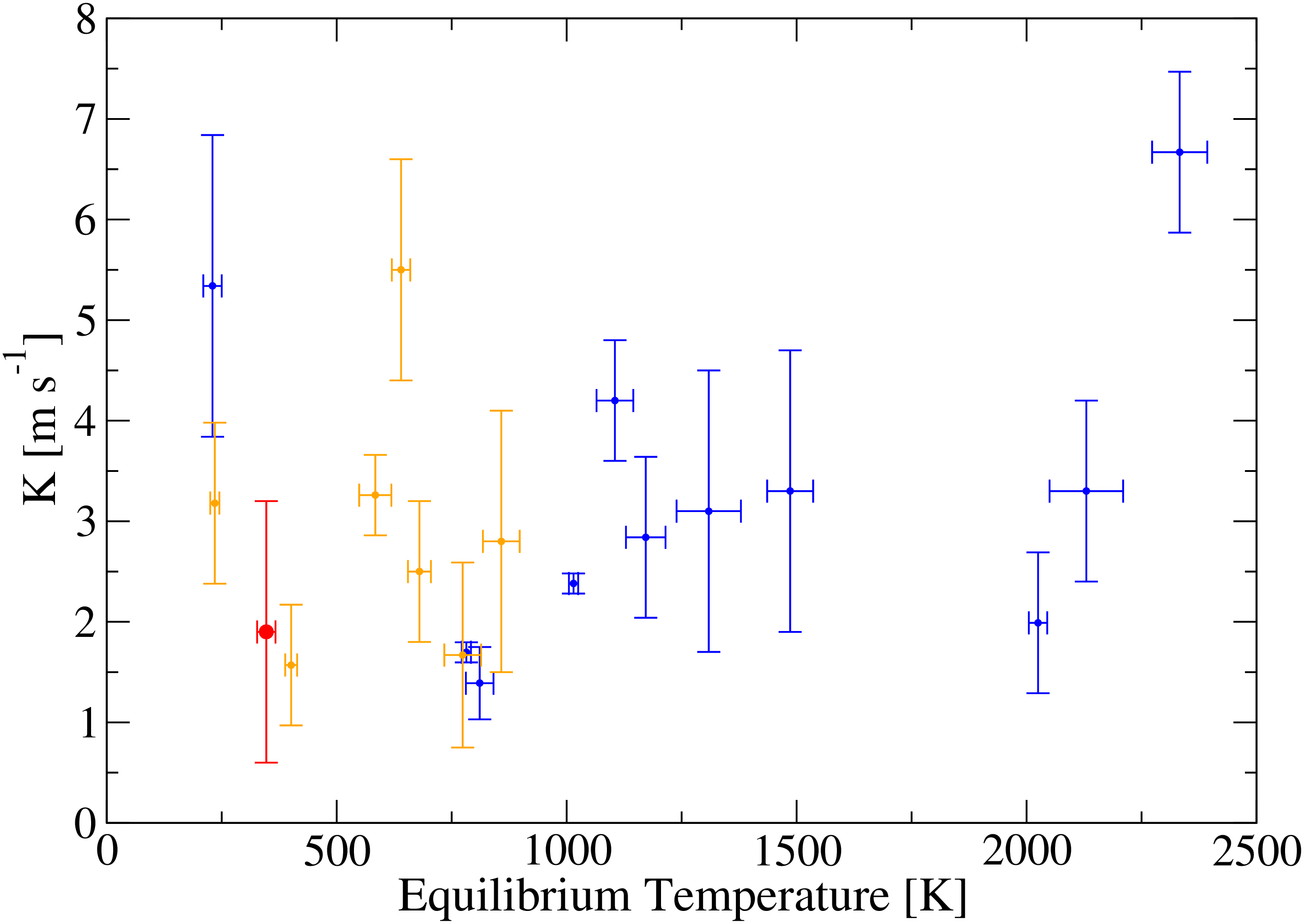}
    \caption{Semi-amplitude of induced radial stellar velocity versus $T_{eq}$ for transiting exoplanets with $R<3$ R$_{\oplus}$ and well measured mass. Blue dots: planets with 1 R$_{\oplus}<R<$ 2 R$_{\oplus}$, orange dots: 2 R$_{\oplus}<R<$ 3 R$_{\oplus}$. Red dot: K2-286b. The semi-amplitude of K2-286b is estimated adopting mass from discussed mass-radius relations.}
    \label{fig:Teqvsvamp}
\end{figure}

\begin{figure}
	\includegraphics [width=0.45\textwidth]{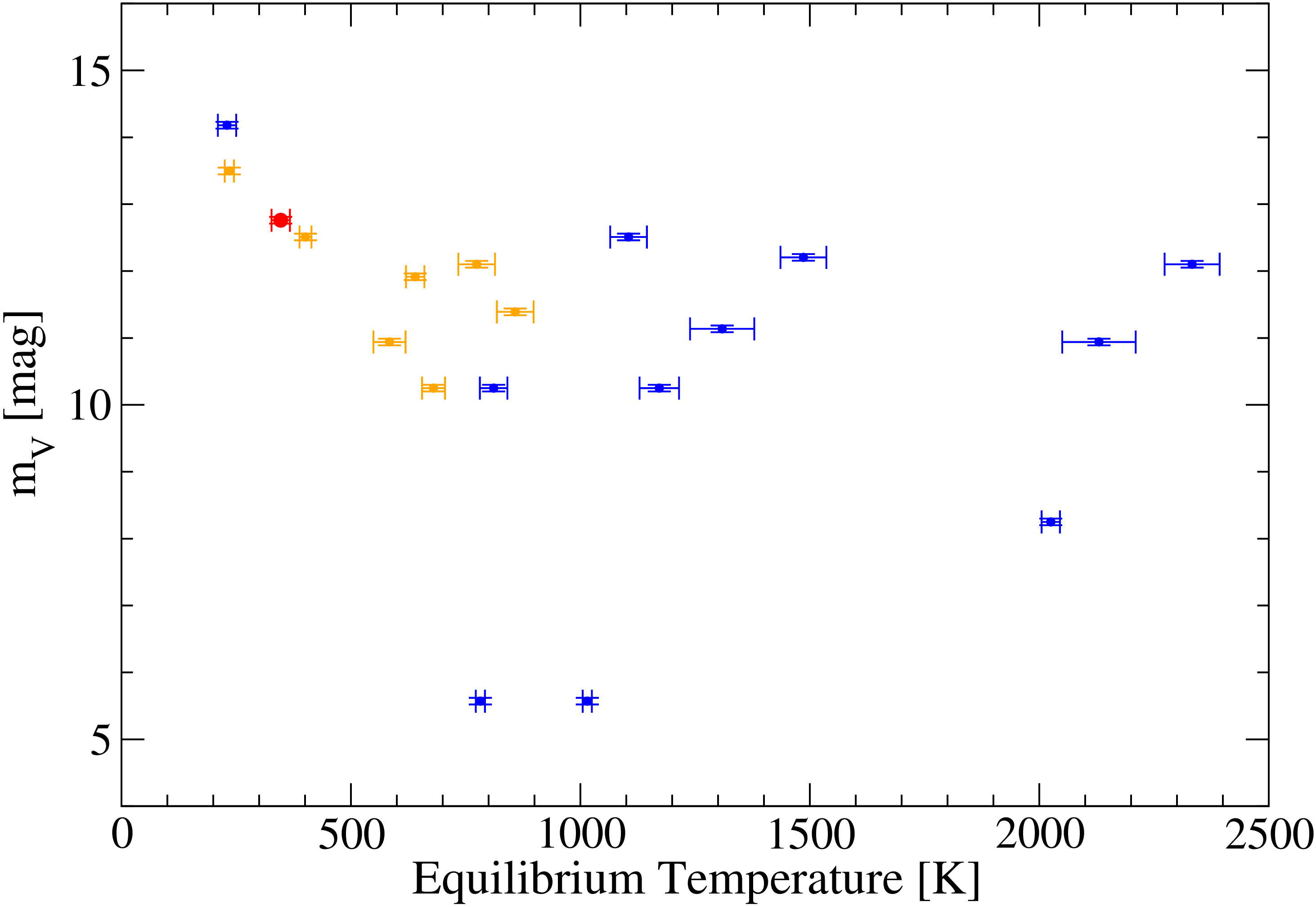}
    \caption{V mag versus $T_{eq}$ for transiting exoplanets with $R<3$ R$_{\oplus}$ and well measured mass. Blue dots: planets with 1 R$_{\oplus}<R<$ 2 R$_{\oplus}$, orange dots: 2 R$_{\oplus}<R<$ 3 R$_{\oplus}$. Red dot: K2-286b.}
    \label{fig:TeqvsmagV}
\end{figure}

In Figures~\ref{fig:Teqvsvamp} $\&$~\ref{fig:TeqvsmagV} we plot $T_{eq}$ against the estimated induced semi amplitudes in stellar velocity variations and the V mag of the host star (respectively) for the same sample of planets as in Fig.~\ref{fig:Teqvsmass}.

\begin{figure}
	\includegraphics [width=0.45\textwidth]{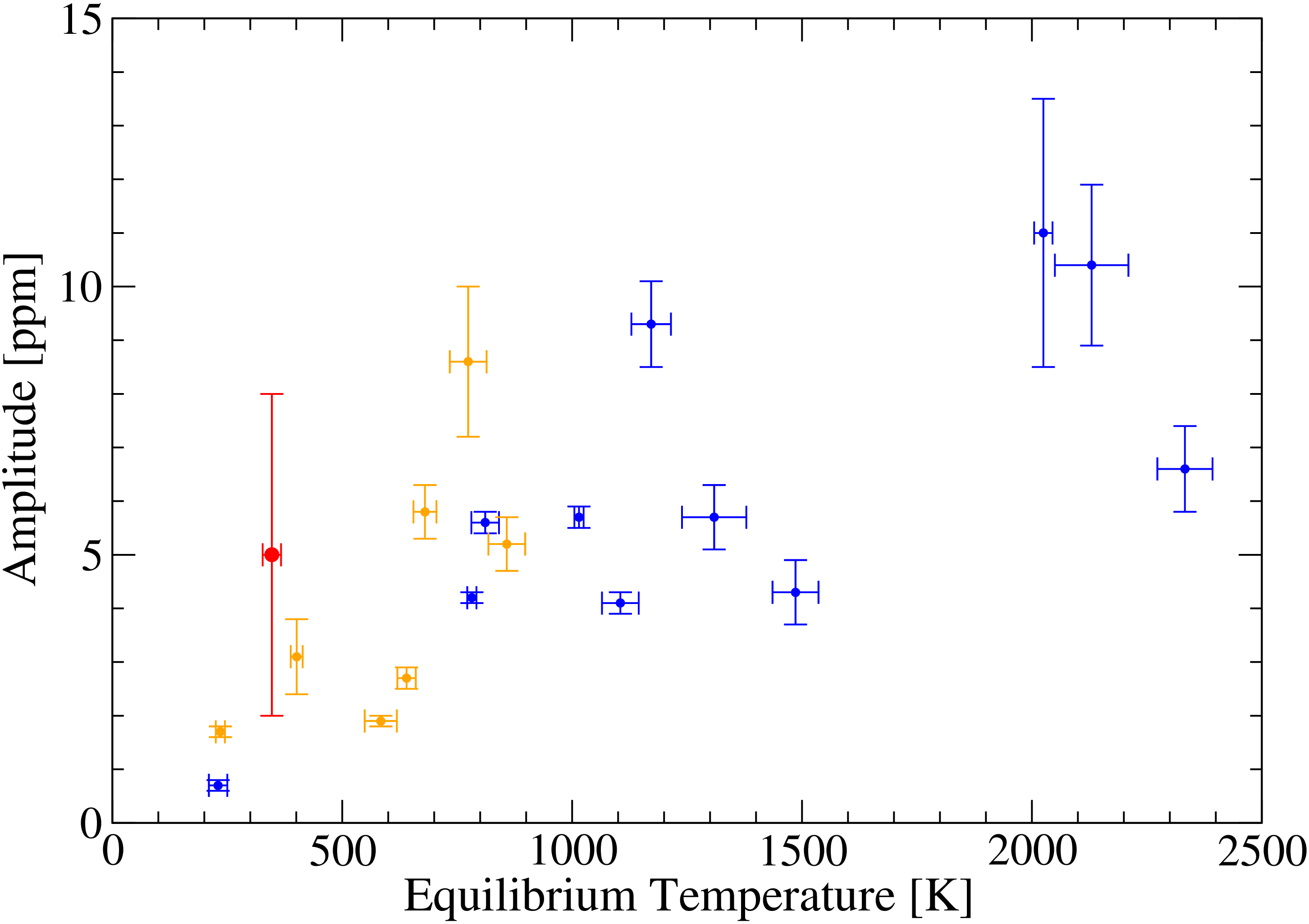}
    \caption{Amplitude of the signal in transit transmission spectroscopy versus $T_{eq}$ for transiting exoplanets with $R<3$ R$_{\oplus}$ and well measured mass. Blue dots: planets with 1 R$_{\oplus}<R<$ 2 R$_{\oplus}$, orange dots: 2 R$_{\oplus}<R<$ 3 R$_{\oplus}$. Red dot: K2-286b. Amplitude of signal for K2-286b is estimated adopting mass from discussed mass-radius relations.}
    \label{fig:TeqvsSNesptrans}
\end{figure}

Understanding the structure of K2-286b will also require transmission spectroscopy observations. The amplitude of the signal in transit transmission spectroscopy can be estimated as $\frac{R_{p} \cdot {h_{\rm eff}}}{(R_{*})^{2}}$ \citep{2016Natur.533..221G} with $h_{\rm eff}$ the effective atmospheric height. $h_{\rm eff}$ is related to the atmospheric scale height H = K$\cdot$T/$\mu$ $\cdot$g (K Boltzmann's constant, T atmospheric temperature, $\mu$ atmospheric mean molecular mass, g surface gravity). Assuming $h_{\rm eff}$ = 7$\cdot$H \citep{2010ApJ...716L..74M} for a transparent volatile dominated atmosphere ($\mu$ = 20) with 0.3 Bond albedo, we estimate the amplitude in transit transmission spectroscopy at $5.0\pm3.0$ ppm. This estimated value is well suited for a future characterization by upcoming facilities such as the James Webb Space Telescope.
In Figure~\ref{fig:TeqvsSNesptrans} we compare the estimated amplitude of the signal in transit transmission spectroscopy of K2-286b with the same sample of transiting small planets with well measured masses. K2-286b has the highest amplitude of signal in the temperate region of the plot, so it is one of the most favourable temperate targets for atmospheric characterization.

\section{Conclusions}
In this work we present the discovery of a super-Earth orbiting close to the inner edge of the habitable zone of the cool dwarf star K2-286.
From acquired spectra we characterized the star as M0V, with $T_{\rm eff}$ = $3926\pm100$ K. We also derived the rest of stellar parameters from empirical relations.

To confirm the planetary origin of the signals found in the light curve of K2-286, we have discarded possible contaminating sources analyzing acquired and archival images, which show no companions or background stars. We have also statistically validated the planetary origin of the signals.

Analyzing the K2 photometric data we find a planetary radius of $2.1\pm0.2$ R$_{\oplus}$, in the upper limit of the 1.5--2.0 R$_{\oplus}$ gap in the radii distribution of exoplanets. $RV$ measurements impose an upper limit to the mass of 22 M$_{\oplus}$, excluding that K2-286b is a gas giant, and from probabilistic mass - radius relations we estimate a planetary mass of $6.8\pm$ 4.3 M$_{\oplus}$.

K2-286b has an estimated equilibrium temperature of $347^{+21}_{-11}$ K. With an orbital period of $27.259\pm0.005$ d and orbital radius of $0.1768^{+0.0175}_{-0.0205}$ AU, K2-286b orbits within the HZ limits according to optimistic models discussed in this work. We have also shown that K2-286 does not exhibit strong stellar activity so K2-286b may experience a more benign environment than other planets orbiting M dwarf stars of later types and might host liquid water depending on its atmospheric conditions. We also have shown that K2-286b is cooler than most of the small exoplanets with well measured masses and radii.

The induced semi-amplitude in stellar velocity variations has been estimated at $1.9^{+1.3}_{-1.2}$ m$\cdot$s$^{-1}$, within the capabilities of high-resolution spectrographs such as ESPRESSO at VLT or EXPRES for the Discovery Channel Telescope, making possible to completely characterize the properties of K2-286b performing radial velocity follow up.

We have also estimated the amplitude of the signal in transit transmission spectroscopy at $5.0\pm3.0$ ppm, within the capabilities of atmospheric characterization by the James Webb Telescope, and shown that K2-286b is one of the most favorable temperate targets for atmospheric characterization.

\section*{Acknowledgements}
EDA, CGG, JCL and JCJ acknowledge Spanish ministry project AYA2017-89121-P. JIGH, BTP, DSA and RRL acknowledge the Spanish ministry project MINECO AYA2014-56359-P and MINECO AYA2017-86389-P. JIGH also acknowledges financial support from the Spanish Ministry of Economy and Competitiveness (MINECO) under the 2013 Ram\'on y Cajal program MINECO RYC-2013-14875. ASM acknowledges financial support from the Swiss National Science Foundation (SNSF).
SLSG and LB acknowledge financial support from the I+D 2015 project
AYA2015- 65887-P (MINECO/FEDER).

Based on observations made with the Gran Telescopio Canarias (GTC), installed in the Spanish Observatorio del Roque de los Muchachos of the Instituto de Astrof\'isica de Canarias, in the island of La Palma.






\bsp	
\label{lastpage}
\end{document}